# Trends in structural and electronic properties for layered SrRu$_2$As$_2$, BaRu$_2$As$_2$, SrRh$_2$As$_2$ and BaRh$_2$As$_2$ from *ab initio* calculations

I. R. Shein* and A. L. Ivanovskii

*Institute of Solid State Chemistry, Ural Branch, Russian Academy of Sciences, Pervomaiskaya St., 91, Ekaterinburg, 620990 Russia*

**A B S T R A C T**

Based on first-principle FLAPW-GGA calculations, we have investigated the systematic trends in structural and electronic properties of a newly discovered group of ThCr$_2$Si$_2$-like arsenides: SrRu$_2$As$_2$, BaRu$_2$As$_2$, SrRh$_2$As$_2$ and BaRh$_2$As$_2$. Our results show that the replacement of an alkaline earth metal (Sr ↔ Ba) and 4*d* metal (Ru ↔ Rh) leads to various types of *anisotropic deformations* of the crystal structure caused by strong anisotropy of inter-atomic bonds.

The band structure, density of states and Fermi surfaces have been evaluated and discussed. Appreciable changes in the near-Fermi bands and the Fermi surface topology found as going from (Sr,Ba)Ru$_2$As$_2$ to (Sr,Ba)Rh$_2$As$_2$ reflect the growth of the 3D-like type of dispersion for these systems, which is accompanied by an increase in the near-Fermi density of states. The inter-atomic bonding in (Sr,Ba)(Ru,Rh)$_2$As$_2$ phases adopts a complex anisotropic character, where the bonding in [(Ru,Rh)$_2$As$_2$] blocks is of a mixed metallic-ionic-covalent type whereas between adjacent [(Ru,Rh)$_2$As$_2$] blocks and (Sr,Ba) atomic sheets, ionic interactions emerge; thus these systems may be classified as *ionic metals*.

* Corresponding author.
*E-mail address:* shein@ihim.uran.ru (I.R. Shein).

## 1. Introduction

The recent discovery [1-7] of new so-called FeAs superconductors (SCs) based on layered pnictides, namely $Ln$FeAsO, $A$FeAsF, $A$Fe$_2$As$_2$ and $A$FeAs (where $Ln$ are rare earth metals such as La, Ce, Sm, Dy, Gd; $A$ are alkaline earth (Ca, Sr, Ba) or alkaline metals (Li, Na)) aroused tremendous interest and stimulated a great research activity of many interesting physical phenomena for these unusual materials, reviews [8,9].

The above FeAs SCs adopt a quasi-two-dimensional crystal structure, where [Fe$_2$As$_2$] blocks are separated either by [$Ln$O], [$A$F] blocks or $A$ atomic sheets. Note that (i) the non-doped parent compounds $Ln$FeAsO and $A$Fe$_2$As$_2$ are located on the border of magnetic instability and commonly exhibit a temperature-dependent structural and magnetic phase transitions with the formation of antiferromagnetic spin ordering; (ii) the electronic bands in the window around the Fermi level are formed mainly by the states of the [Fe$_2$As$_2$] blocks and play the main role in superconductivity whereas the [$Ln$O], [$A$F] blocks ($A$ atomic sheets) serve as "charge reservoirs", and (iii) superconductivity emerges either by hole or electron doping of the parent compounds, see [1-9].

In particular, a broad family of the so-called FeAs "122" superconductors (with transition temperatures to $T_C$ = 38K, see [8-10]) based on $A$Fe$_2$As$_2$ phases ($A$ = Ca, Sr or Ba) was prepared by: (i) hole doping, *i.e.* by partial substitution of alkaline metals for alkaline earth metals or (ii) by partial replacement of Fe (in [Fe$_2$As$_2$] blocks) by other 3$d$ transition metals, $TM$ = Mn, Co or Ni. Note that all these materials have a high content of **magnetic metals** (Fe, Mn, Co and Ni). Interestingly, the same result may be achieved by **partial substitution** of some **non-magnetic 4d, 5d metals** (Ru, Ir *etc*.) for **magnetic 3d metal** (Fe), see [10-12].

In view of these circumstances, the recent reports on the synthesis of a new group of ThCr$_2$Si$_2$-like iron-free arsenides, namely, SrRu$_2$As$_2$, BaRu$_2$As$_2$ [13] SrRh$_2$As$_2$ [14] and BaRh$_2$As$_2$ [15] are very interesting. Really, (i) these materials are isostructural to the above family of "122" FeAs SCs, (ii) SrRu$_2$As$_2$, BaRu$_2$As$_2$ are also isoelectronic to SrFe$_2$As$_2$, BaFe$_2$As$_2$ - the well known parent phases for "122" FeAs SCs, whereas SrRh$_2$As$_2$ and BaRh$_2$As$_2$ have two additional electrons, but (iii) for these phases the *magnetic metal* (Fe) *is completely replaced by the non-magnetic metals* (Ru, Rh).

In this Communication, we studied the structural, electronic properties and intra-atomic bonding for the newly discovered ThCr$_2$Si$_2$-like arsenides SrRu$_2$As$_2$, BaRu$_2$As$_2$, SrRh$_2$As$_2$ and BaRh$_2$As$_2$ by means of the first-principle FLAPW-GGA method with the purpose to evaluate the systematic trends of the above properties as a function of the alkaline earth metal type (Sr *versus* Ba, *i.e.* SrRu$_2$As$_2$ ↔ BaRu$_2$As$_2$ and SrRh$_2$As$_2$ ↔ BaRh$_2$As$_2$) and the 4*d TM* type (Ru *versus* Rh, *i.e.* SrRu$_2$As$_2$ ↔ SrRh$_2$As$_2$ and BaRu$_2$P$_2$ ↔ BaRh$_2$As$_2$) - in comparison with BaFe$_2$As$_2$ as a parent phase for "122" FeAs SCs. To our knowledge, the electronic properties of these compounds have not been investigated before (except BaRh$_2$As$_2$ [15]).

**2. Computational aspects**

All the considered ternary arsenides $AM_2$As$_2$ ($A$= Sr, Ba; $M$ = Ru, Rh) crystallize in the quasi-two-dimensional ThCr$_2$Si$_2$-type tetragonal structure, space group I4/*mmm*; Z=2. The structure is built up of [$M_2$As$_2$] blocks alternating with $A$ atomic sheets stacked along the *z* axis, as depicted in Fig. 1. The atomic positions are: $A$: 2*a* (0,0,0), $M$: 4*d* (½,0,½) and As atoms: 4*e* (0,0,$z_{As}$), where $z_{As}$ are the so-called internal coordinates governing the $M$-As distances and the distortion of the $M$As$_4$ tetrahedra around the $M$ in the [$M_2$As$_2$] blocks.

Our calculations were carried out by means of the full-potential method with mixed basis APW+lo (LAPW) implemented in the WIEN2k suite of programs [16]. The generalized gradient correction (GGA) to exchange-correlation potential in the PBE form [17] was used. The plane-wave expansion was taken to $R_{MT} \times K_{MAX}$ equal to 7, and the $k$ sampling with 10×10×10 $k$-points in the Brillouin zone was used. The calculations were performed with full-lattice optimization including internal $z_{As}$ coordinates. The self-consistent calculations were considered to be converged when the difference in the total energy of the crystal did not exceed 0.1 mRy and the difference in the total electronic charge did not exceed 0.001 $e$ as calculated at consecutive steps.

## 3. Results and discussion

At the first step, the equilibrium structural parameters for the arsenides $AM_2As_2$ are determined; the calculated values are presented in Table 1 and are in reasonable agreement with the available experiments [13-15]. Some deviations of our results from the experimental data for SrRh$_2$As$_2$ may be connected with the circumstance that this arsenide has a set of polymorphs (α-γ), and the examined sample [14] of the high-temperature polymorph of ThCr$_2$Si$_2$-type can contain inclusions of other polymorphs. This assumption is supported by the recent discovery of structural domains near the temperature region of transition from tetragonal to orthorhombic phase [18].

Our results show that replacements of the alkaline earth metal (Sr ↔ Ba) or 4$d$ metal (Ru ↔ Rh) lead to *anisotropic deformations* of the crystal structure caused by strong anisotropy of inter-atomic bonds (see also [19,20]), but the type of such deformations may be quite different. Really, when going from SrRu$_2$As$_2$ to BaRu$_2$As$_2$, *i.e.* when a small Sr atom (atomic radius $R^{at}$ = 2.15 Å) is replaced by a larger Ba atom ($R^{at}$ = 2.21 Å) inside $A$ sheets, the inter-layer distance grows appreciably (parameter $c$, by about 1.02 Å) whereas the parameter $a$ slightly decreases (~ 0.01 Å). On the contrary, when going from SrRh$_2$As$_2$ to BaRh$_2$As$_2$, both parameters ($c$ and $a$) grow. On the other hand, when a Ru atom is replaced by a Rh atom (with very close atomic radii, $R^{at}$ ~ 1.34 Å), *i.e.* for the series $SrRu_2As_2$ ↔ $SrRh_2As_2$ and $BaRu_2P_2$ ↔ $BaRh_2As_2$, the parameter *a* decreases, whereas the parameter *c* grows. Finally, in the "mixed" case, i.e. when a Sr atom is replaced by a Ba atom and a Ru atom is replaced by a Rh atom ($SrRu_2As_2$ ↔ $BaRh_2As_2$), the parameter *c* grows, whereas the parameter *a* decreases. The corresponding changes of internal coordinates ($z_{As}$), inter-atomic distances and angles are summarized in Table 1. Thus, the results obtained reveal that various atomic replacements inside "122" phases can lead to essential anisotropic changes of their structural properties and are favorable for fine tuning of their geometry - in particular, in view of sensitivity of superconductivity to structural changes, see [21,22].

Figures 2 and 3 show the band structures and total and atomic-resolved *l*-projected DOSs in $SrRu_2As_2$, $BaRu_2As_2$, $SrRh_2As_2$ and $BaRh_2As_2$ phases as calculated for equilibrium geometries.

The common features of the band structure for these phases will be illustrated using $SrRu_2As_2$ as an example. Here the two lowest bands lying around -11 eV below the Fermi level ($E_F$) arise mainly from As 4*s* states and are separated by a gap from the near-Fermi valence bands, which are located in the energy range from -5.9 eV to $E_F$ and are formed predominantly by Ru 4*d* and As 4*p* states. The corresponding DOS include two subband B and C, Fig. 3. The subband B contains strongly hybridized Ru 4*d* - As 4*p* states, which are responsible for the covalent Ru-As bonding. The intense peak C in the DOS is due to the Ru 4*d*-like bands with low E(*k*) dispersion, which are located around -2 eV, and these states participate in metallic-like Ru-Ru bonds. Finally, the bottom of the conduction band (subband D) is also made up basically of Ru 4*d* states with an admixture of anti-bonding As 4*p* states. Thus, the near-Fermi region is formed mainly by the states of [$Ru_2As_2$] blocks. Besides, it is noteworthy that the contributions from the valence states of Sr to the occupied subbands are negligible, *i.e.* in SrRu$_2$As$_2$ (as well as in other examined arsenides) the alkaline earth atoms are in the cation form close to $A^{2+}$.

The electronic band structure of the ThCr$_2$Si$_2$-like iron arsenides $A$Fe$_2$As$_2$ has been described frequently, see [8,9,23]. The results show that the Fermi surface (FS) consists of electron-like pockets at the corners (M) and typical hole-like cylinders around Γ, which reflect the two-dimensional-like dispersion for these "122" FeAs systems. This FS nesting is unstable and is favorable for charge- or spin-density ordering, often connected with structural distortions [24]. Our calculations predict clear differences in FSs for isostructural and isoelectronic SrRu$_2$As$_2$ and BaRu$_2$As$_2$ (Fig. 4) when the above cylinders around Γ are split into a pair of disconnected closed pockets centered at Z. The general conclusion is that the hole sheets change from two-dimensional (2D) to three-dimensional (3D) dispersion when going from $A$Fe$_2$As$_2$ to SrRu$_2$As$_2$ and BaRu$_2$As$_2$. Our data show also that the reconstruction of the 2D-like FS topology for SrRu$_2$As$_2$ and BaRu$_2$As$_2$ (as one of the necessary conditions of superconductivity for these systems) may be due to partial "emptying" of the valence zone, when the hole sheets appear at Γ, see Fig. 2. It can be achieved, for example, by introduction of alkaline metal ions into $A$ sheets.

Let us discuss the most important differences in Ru- and Rh-based "122" phases, focusing on the near-Fermi region, Figs. 2-4 and Table 2. The arsenides SrRh$_2$As$_2$, BaRh$_2$As$_2$ have an increased valence electron count (+2 $e$) as compared with their Ru-based counterparts. As a result, $E_F$ moves from the DOS minimum. In the case of SrRh$_2$As$_2$ and BaRh$_2$As$_2$, also the contributions from the Rh 4$d$ band (see Table 2) clearly dominate in the vicinity of the Fermi level, but appreciable changes of the near-Fermi bands and the Fermi surface topology have been found, which reflect the 3D-like type of dispersion for these systems.

According to our estimations, the total DOSs at the Fermi level, N($E_F$), for the examined materials increase in the sequence:

$$\text{SrRu}_2\text{As}_2 < \text{BaRu}_2\text{As}_2 < \text{SrRh}_2\text{As}_2 < \text{BaRh}_2\text{As}_2 \qquad (1)$$

The obtained data also allow us to estimate the Sommerfeld constants ($\gamma$) and the Pauli paramagnetic susceptibility ($\chi$) for these arsenides under the assumption of the free electron model as $\gamma = (\pi^2/3)N(E_F)k^2_B$ and $\chi = \mu_B^2 N(E_F)$. It is seen from Table 2 that the calculated $\gamma$ values agree well with available experimental data, and both $\gamma$ and $\chi$ decrease in the sequence (1). The only exception is the phase $\text{BaRh}_2\text{As}_2$, for which the experimental value of $\gamma^{exp}$ is twice smaller (4.8 mJ·K$^{-2}$·mol$^{-1}$) [15] than our estimation $\gamma^{theor}$ = 8.61 mJ·K$^{-2}$·mol$^{-1}$ (a similar value is also obtained from the LDA calculations [15] $\gamma^{theor}$ = 8.23 mJ·K$^{-2}$·mol$^{-1}$).

One of the reasons of this divergence may be strong electron correlations. Really, using experimental values $\gamma^{exp}$ and $\chi^{exp}$ [15] we can estimate the Wilson ratio $R_W = \chi/\gamma \times \pi^2 k_B^2/3\mu_B^2$ to be close to 1.7. The fact that many of strongly correlated Fermi liquids show a Wilson ration close to 2 is indicative of relatively strong electron correlations. On the other hand, the same estimations using the experimental data [13] for related systems yield the values $R_W$ = 0.96 ($\text{SrRu}_2\text{As}_2$), 1.2 ($\text{BaRu}_2\text{As}_2$) and 0.98 ($\text{BaRh}_2\text{P}$ [25]), which imply no strong electron correlations and magnetic fluctuations. This point requires additional studies.

To describe the intra-atomic bonding for the examined $AM_2\text{As}_2$, it is convenient to begin with a standard ionic picture, which considers the usual oxidation numbers of atoms: $A^{2+}$, $M^{2+}$ and $\text{As}^{3-}$. Then, the charge states of the blocks are $[A]^{2+}$ and $[(M^{2+})_2(\text{As}^{3-})_2]^{2-}$, i.e. the charge transfer occurs from $A^{2+}$ sheets to $[M_2\text{As}_2]^{2-}$ blocks. Besides, inside the $[M_2\text{As}_2]$ blocks, the ionic bonding takes place between $M$-As atoms. The character of covalent bonding, i.e. the formation of $M$-As bonds owing to hybridization of $M$ 4$d$ - As 4$p$ states, is clearly visible from site-projected DOSs, Fig. 3. In addition, inside the $[M_2\text{As}_2]$ blocks the metallic-like $M$-$M$ bonding occurs due to overlapping of the near-Fermi $M$ 4$d$ states. The described picture is shown in Fig. 5, where the charge density map for $\text{SrRu}_2\text{As}_2$ is presented. Finally, let us note that no As-As bonds are present between the adjacent $[M_2As_2]/[M_2As_2]$ blocks (Fig. 5) - in contrast to $AFe_2As_2$ materials. This may be explained by the differences in the strength of (Ru,Rh)-As and Fe-As bonds, which accompany the formation of As-As interactions [26].

Thus, summarizing the above results, the intra-atomic bonding for $AM_2As_2$ phases can be classified as a high-anisotropic mixture of ionic, covalent and metallic contributions, where inside the $[M_2As_2]$ blocks, mixed covalent-ionic-metallic bonds $M$-As take place (owing to hybridization of M $4d$ – As $4p$ states, M → As charge transfer and delocalized near-Fermi M $4d$ states, respectively), whereas between the adjacent $[M_2As_2]$ blocks and $A$ atomic sheets, ionic bonds emerge owing to [NiP(As)] $A \to [M_2As_2]$ charge transfer.

## 4. Conclusions

In summary, by means of the FLAPW-GGA approach, we have systematically studied the structural and electronic properties of a group of newly synthesized tetragonal ternary arsenides $SrRu_2As_2$, $BaRu_2As_2$, $SrRh_2As_2$ and $BaRh_2As_2$.

Our results show that replacements of alkali earth metal (Sr ↔ Ba) or $4d$ metal (Ru ↔ Rh) lead to various types of *anisotropic deformations* of the crystals structure caused by strong anisotropy of inter-atomic bonds.

The basic band structure pictures of all four Ru,Rh-based "122" phases are similar, and the near-Fermi valence bands in these phases are derived from the (Ru,Rh) $4d$ states with an admixture of the As $4p$ states. Nevertheless some important differences take place owing to increased valence electron count as going from Ru-based to Rh-based "122" phases. In particular, appreciable changes in the near-Fermi bands and the Fermi surface topology have been found, which reflect the growth of the 3D-like type of dispersion for these systems, which is accompanied by an increase of the near-Fermi density of states.

Our analysis reveals also that the intra-atomic bonding for $AM_2As_2$ has a high-anisotropic character, where inside the $[M_2As_2]$ blocks, mixed covalent-ionic-metallic bonds take places, whereas between the adjacent $[M_2As_2]$ blocks and $A$ atomic sheets, ionic bonds emerge. Thus, the examined phases may be classified as *ionic metals*.

**Acknowledgments**

Financial support from the RFBR (Grant 09-03-00946-a) is gratefully acknowledged.

Table 1.
The optimized lattice parameters (*a* and *c*, in Å), internal coordinates ($z_{As}$), some inter-atomic distances (*d*, in Å) and angles As-(Ru,Rh)-As (θ, in deg.) for ternary arsenides $SrRu_2As_2$, $BaRu_2As_2$, $SrRh_2As_2$ and $BaRh_2As_2$.

| phase/parameter | $SrRu_2As_2$ | $BaRu_2As_2$ | $SrRh_2As_2$ | $BaRh_2As_2$ |
|---|---|---|---|---|
| *a* | 4.2068 (4.1713 [13]) | 4.1925 (4.1525 [13]) | 4.0801 (4.112 [14]) | 4.0894 (4.0565 [15]; 4.053 [14]) |
| *c* | 11.2903 (11.1845[13]) | 12.3136 (12.2504 [13]) | 12.1990 (11.432 [14]) | 12.8423 (12.797 [15]; 12.770 [14]) |
| *c/a* | 2.6838 (2.6813 [13] ) | 2.9371 ( 2.9501[13]) | 2.9901 (2.7802 [14]) | 3.1404 (3.1547 [15]; 3.1508 [14]) |
| $z_{As}$ | 0.3591 (0.3612 [13]) | 0.3510 (0.3527 [13]) | 0.3647 (0.3641 [14]) | 0.3578 (0.3566 [15]; 03569 [14]) |
| $d^{Ru-Ru\ (Rh-Rh)}$ | 2.975 | 2.966 | 2.885 (2.908 [14]) | 2.892 (2.866 [14]) |
| $d^{Ru-As\ (Rh-As)}$ | 2.437 | 2.438 | 2.476 (2.435 [14]) | 2.469 (2.443 [14]) |
| $d^{As-As}$ | 3.182 | 3.671 | 3.302 (3.109 [14]) | 3.653 (3.655 [14]) |
| θ | 119.3 (118.4 [13]) | 118.7 (117.6 [13]) | 111.1 (106.7-115.2 [14]) | 111.8 (108.2-112.1 [14]) |

* available experimental data are given in parentheses.

Table 2. Total and partial densities of states at the Fermi level (in states/eV·f.u.), electronic heat capacity γ (in mJ·K$^{-2}$·mol$^{-1}$) and molar Pauli paramagnetic susceptibility χ (in 10$^{-4}$ emu/mol) for ternary arsenides SrRu$_2$As$_2$, BaRu$_2$As$_2$, SrRh$_2$As$_2$ and BaRh$_2$As$_2$.

| system/parameter | As 4$p$ | (Ru,Rh) 4$d$ | total | γ | χ |
|---|---|---|---|---|---|
| SrRu$_2$As$_2$ | 0.155 | 0.802 | 1.708 | 4.03 | 0.550 |
|  |  |  |  | (4.1 [13]) | (0.550 [13]) |
| BaRu$_2$As$_2$ | 0.113 | 0.864 | 1.713 | 4.04 | 0.551 |
|  |  |  |  | (4.9 [13]) | (0.680 [13]) |
| SrRh$_2$As$_2$ | 0.284 | 1.716 | 3.399 | 8.01 | 1.09 |
| BaRh$_2$As$_2$ | 0.266 | 1.664 | 3.651 | 8.61 | 1.18 |
|  | (0.41** [15]) | (1.54** [15]) | (3.49** [15]) | (4.8 [15]) | (1.12 [15]) |

\* available experimental data are given in parentheses.
\*\* theoretical data [15]

**FIGURES**

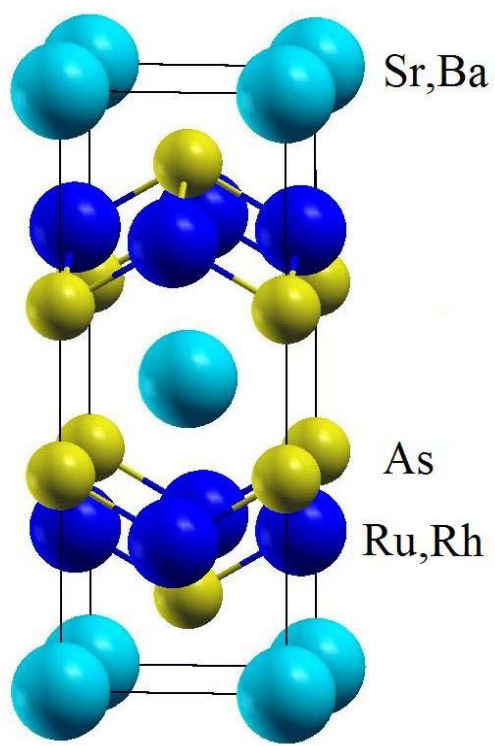

Fig. 1. (*Color online*) Crystal structure of tetragonal arsenides (Sr,Ba)(Ru,Rh)$_2$As$_2$.

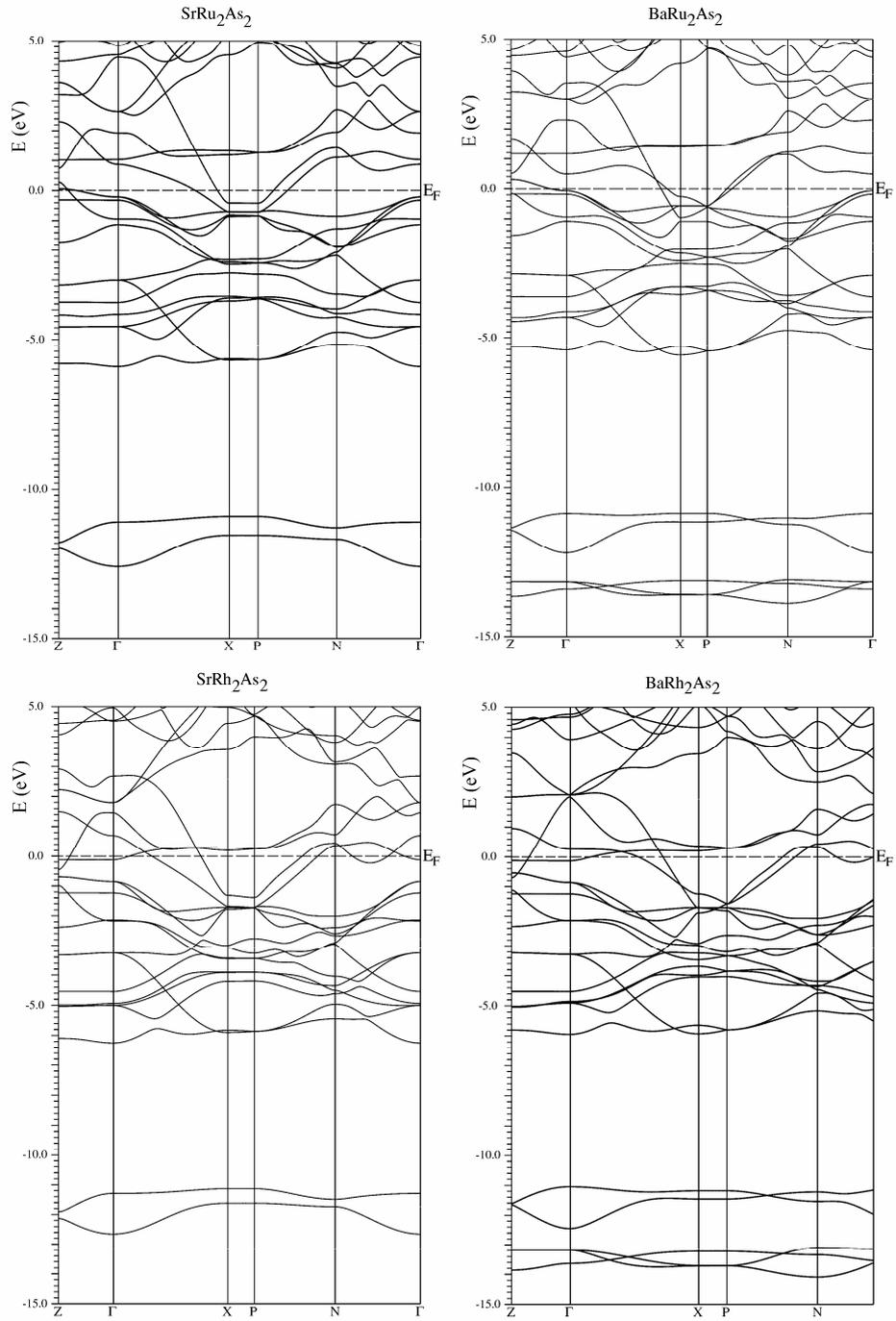

Fig. 2. Electronic band structures of SrRu$_2$As$_2$ (1), BaRu$_2$As$_2$ (2), SrRh$_2$As$_2$ (3) and BaRh$_2$As$_2$ (4).

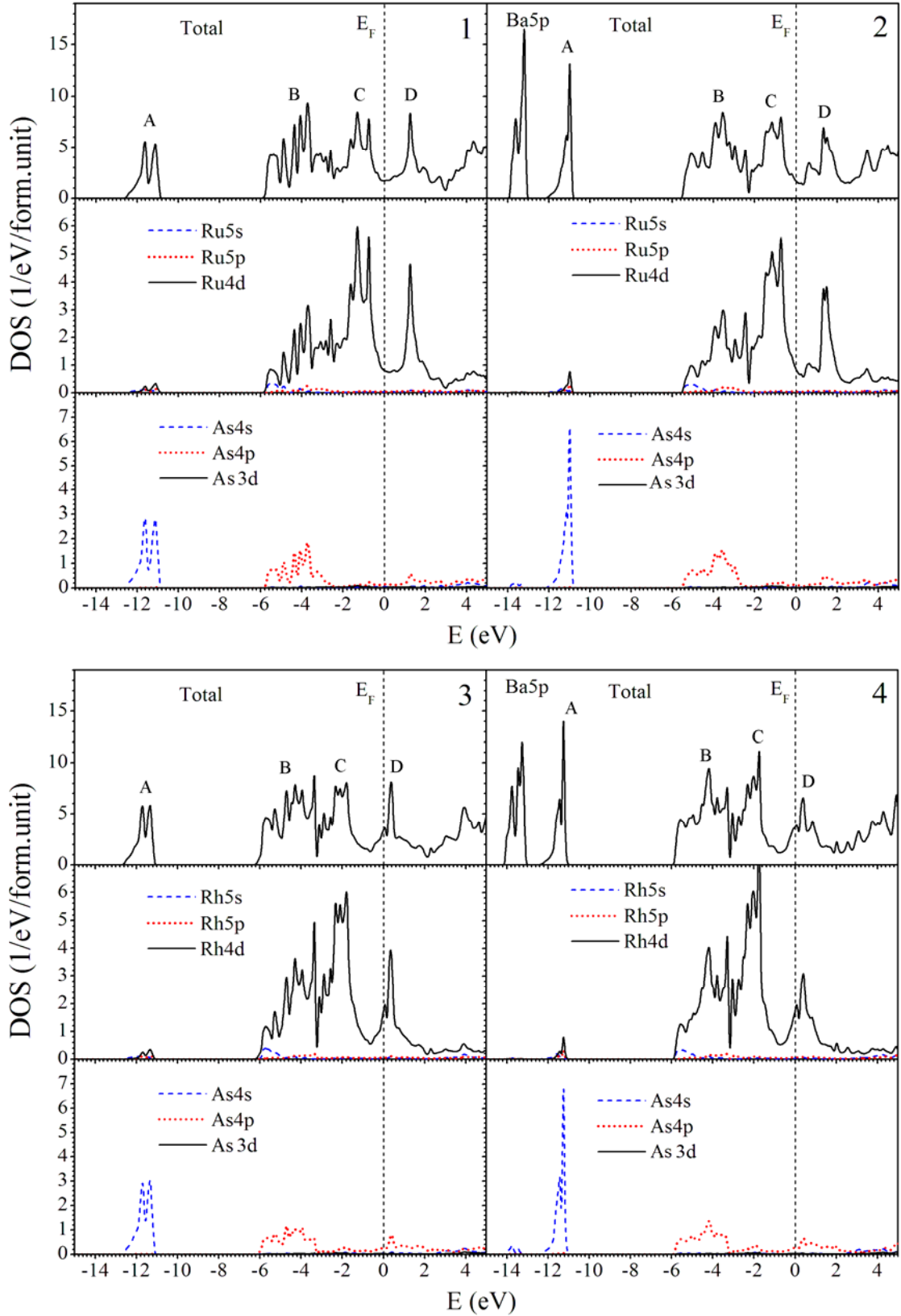

Fig. 3. (*Color online*) Total and partial densities of states of SrRu$_2$As$_2$ (1), BaRu$_2$As$_2$ (2), SrRh$_2$As$_2$ (3) and BaRh$_2$As$_2$ (4).

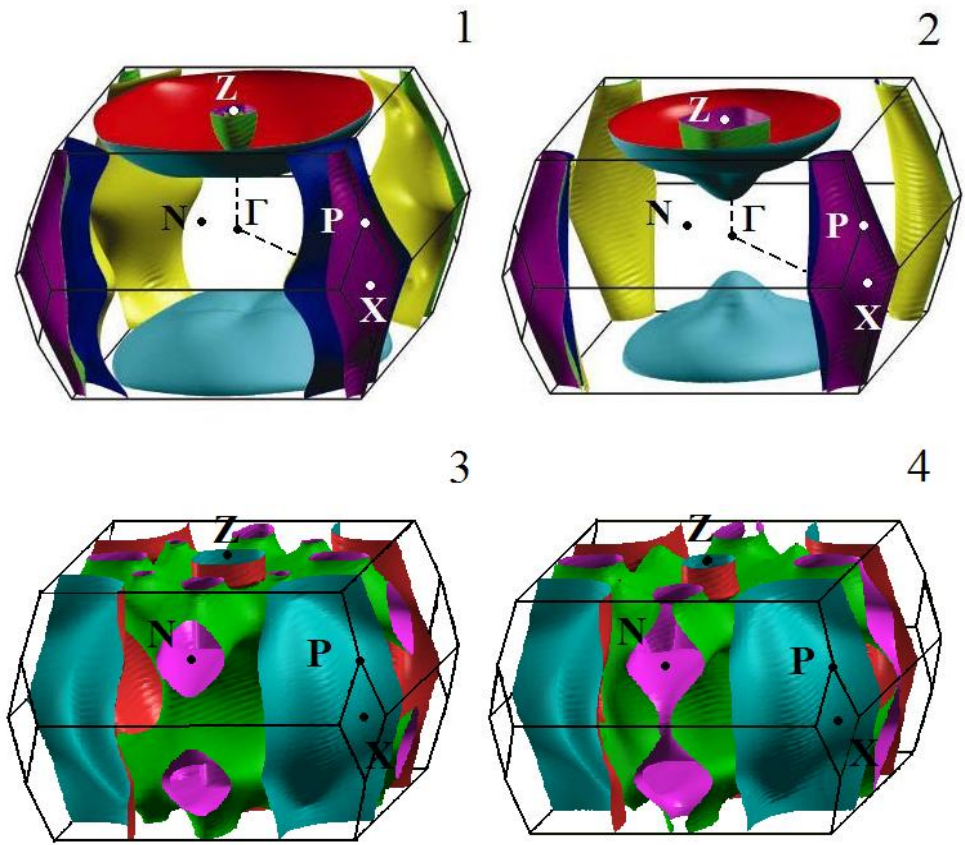

Fig. 4. (*Color online*) The Fermi surfaces of SrRu$_2$As$_2$ (1), BaRu$_2$As$_2$ (2), SrRh$_2$As$_2$ (3) and BaRh$_2$As$_2$ (4).

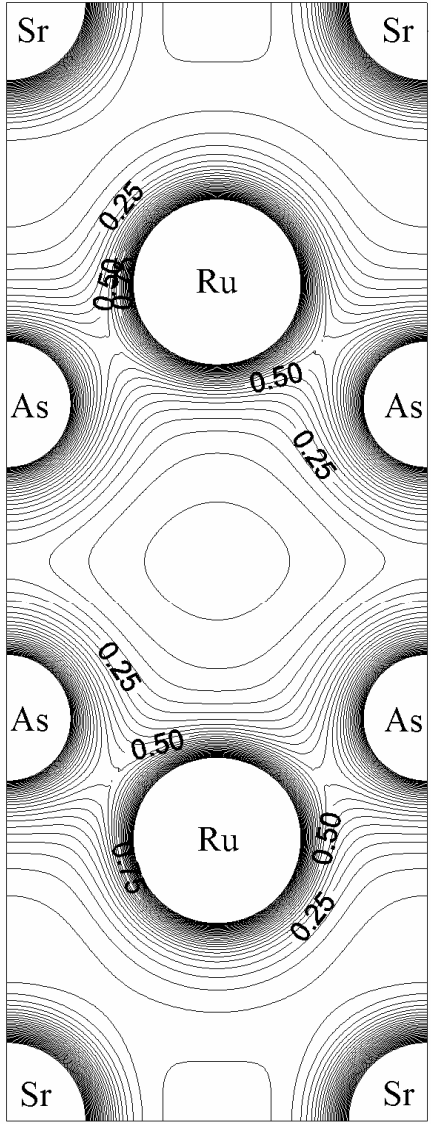

Fig. 5. Valence charge density maps (in e/Å$^3$) in [100] plane for SrRu$_2$As$_2$.